\documentclass[a4paper,twocolumn,12pt]{quantumarticle}
\pdfoutput=1
\usepackage[numbers]{natbib}
\usepackage{yquant}
\usepackage[USenglish]{babel}

\usepackage[utf8]{inputenc}
\usepackage{enumitem}
\usepackage{makecell}
\usepackage{subfig}
\usepackage{tikz,hologo,xspace,pbox}
\usepackage{lipsum}

\usepackage{float}
\makeatletter
\let\newfloat\newfloat@ltx
\makeatother
\usepackage{algorithm}

\usepackage{algorithmic}

\usepackage{geometry} 
\usepackage{graphicx} 

\usepackage{booktabs}
\usepackage{array} 
\usepackage{paralist} 
\usepackage{verbatim} 
\usepackage{subfig} 
\usepackage{amsmath}
\usepackage{amssymb}

\usepackage{fancyvrb}

\usepackage{fancyhdr} 
\title{
Automated Generation of Shuttling Sequences for a Linear Segmented Ion Trap Quantum Computer}
\author[2]{Jonathan Durandau}
\author[1]{Janis Wagner}
\author[2]{Frédéric Mailhot}
\author[2]{Charles-Antoine Brunet}
\author[1]{Ferdinand Schmidt-Kaler}
\author[1]{Ulrich Poschinger}
\author[2]{Yves Bérubé-Lauzière}

\affil[1]{QUANTUM, Institute of Physics, Johannes Gutenberg University, Staudingerweg 7, 55128 Mainz, Germany}
\affil[2]{Institut Quantique and Département de génie électrique et de génie informatique, Université de Sherbrooke, Sherbrooke, Québec, J1K 2R1, Canada}
\begin{document}
\maketitle

\begin{abstract}

A promising approach for scaling-up trapped-ion quantum computer platforms is by storing multiple trapped-ion qubit sets (“ion crystals”) in segmented microchip traps and to interconnect these via physical movement of the ions (“shuttling”). Already for realizing quantum circuits with moderate complexity, the design of suitable qubit assignments and shuttling schedules require automation.
Here, we describe and test algorithms which address exactly these tasks. We describe an algorithm for fully automated generation of shuttling schedules, complying to constraints imposed by a given trap structure.
Furthermore, we introduce different methods for initial qubit assignment and compare these for random circuit (of up to 20 qubits) and quantum Fourier transform-like circuits, and generalized Toffoli gates of up to 40 qubits each. We find that for quantum circuits which contain a fixed structure, advanced assignment algorithms can serve to reduce the shuttling overhead.

\end{abstract}

\section*{Introduction} 

Quantum computing platforms currently undergo a maturation process towards becoming robust programmable devices, allowing for first applications in the noisy intermediate-scale quantum computing (NISQ) regime~\cite{Preskill2018NISQEra}, operating on register sizes of the order of 10 to 100 qubits.
Among the leading qubit realizations are atomic ions confined in radio frequency traps, with the key advantages of long coherence times along with high-fidelity gate and readout operations.

One trapped-ion-based architecture which has seen substantial progress throughout the past decades operates on static linear ion strings - referred to as \textit{ion crystals} in the following - \cite{Figgatt2019,Schindler2013}, and employs addressed laser beams for driving gate operations on specific subsets of qubits.
For this approach, the degree of control - in terms of e.g. gate fidelities - decreases with increasing register size. To achieve scalability beyond maximum linear ion string register sizes of some tens of qubits, the \textit{Quantum CCD} (QCCD) architecture has been proposed \cite{Kielpinski2002ArchitectureFA}.
Here, the ion qubits are confined in a \textit{segmented} ion trap, \textit{i.e.}
a complex micro-structured electrode array. The qubits are stored in small subsets, and gate operations are carried out locally on these, at one or more fixed sites of the trap. With this approach, control imperfections scale less detrimentally with the register size. Between consecutive gate operations, the register is dynamically rearranged via \textit{shuttling operations}, \textit{i.e.} qubit ions are moved within the trap structure via the application of suitable voltage ramps to the trap electrodes, in analogy to moving charges on imaging CCD chips. Recent realizations of such QCCD platforms are described in \cite{shuttlingIonTrapQuantum,Pino2021,Hilder2022}.

As hardware capabilities of quantum computing platforms increase, so do the requirements on the control software to generate sequences of hardware-level commands for executing a given quantum algorithm. For the specific case of the QCCD architecture, this breaks down into two main stages:
First, a quantum circuit needs to be compiled into the set of native gate operations allowed by the platform \cite{Kreppel2022, Schmale2022}. Second, another compilation step is required to generate a \textit{shuttling schedule}. This schedule consists of a sequence of storage configurations, each configuration yielding the required qubit subsets to be stored at the manipulation site(s) and idle qubits at different, well-defined storage sites. Between two consecutive storage configurations, the register is to be dynamically rearranged via shuttling operations. Shuttling schedules have to be computed in an automated, efficient, and robust manner, minimizing the overhead of shuttling operations, and taking into account the specific hardware constraints of a given platform. This is the problem tackled in the present work.

The problem of generating shuttling schedules has been addressed recently in \cite{winnie2020,Saki2021} in the context of an envisioned large-scale platform consisting of an array of identical traps, and a fixed algorithm-independent shuttling protocol. In \cite{wu2020tilt}, the generation of optimal shuttling schedules for an architecture based on a long monolithic linear ion string is described. In that architecture, the entire ion string is to be moved through a laser addressing zone, in which multiple laser beams simultaneously manipulate multiple neighboring ion qubits. The present work describes a complete algorithmic framework for generating algorithm-dependent shuttling schedules in the limit where the register is broken down into subsets of two qubits at most, and the reconfiguration operations consist of different types of shuttling operations.

The outline of this manuscript is as follows: Sect.~\ref{sect:ProbStatmnt} introduces the linear segmented ion trap quantum computer architecture, and outlines the problem of generating a viable ion shuttling schedule. Constraints and cost evaluation are then discussed in Sect.~\ref{sect:ConstraintsCosts}. The overall approach for generating a shuttling sequence is presented in Sect.~\ref{sect:OverallApproach}. Sect.~\ref{sect:InitialIonOrder} and~\ref{sect:Shuttling} respectively present algorithms for establishing the initial ion ordering and subsequent shuttling. Implementation details of the overall approach are presented in Sect.~\ref{sect:Implement}, and example results are provided in Sect.~\ref{sect:Results}. A proof of optimality follows in Sect.~\ref{sect:DiscussProof} along with a discussion.

\begin{figure*}[htp!]
    \centering
\centering
 	\includegraphics[width=0.7\textwidth,trim={0 2.5cm 0 2.5cm},clip]{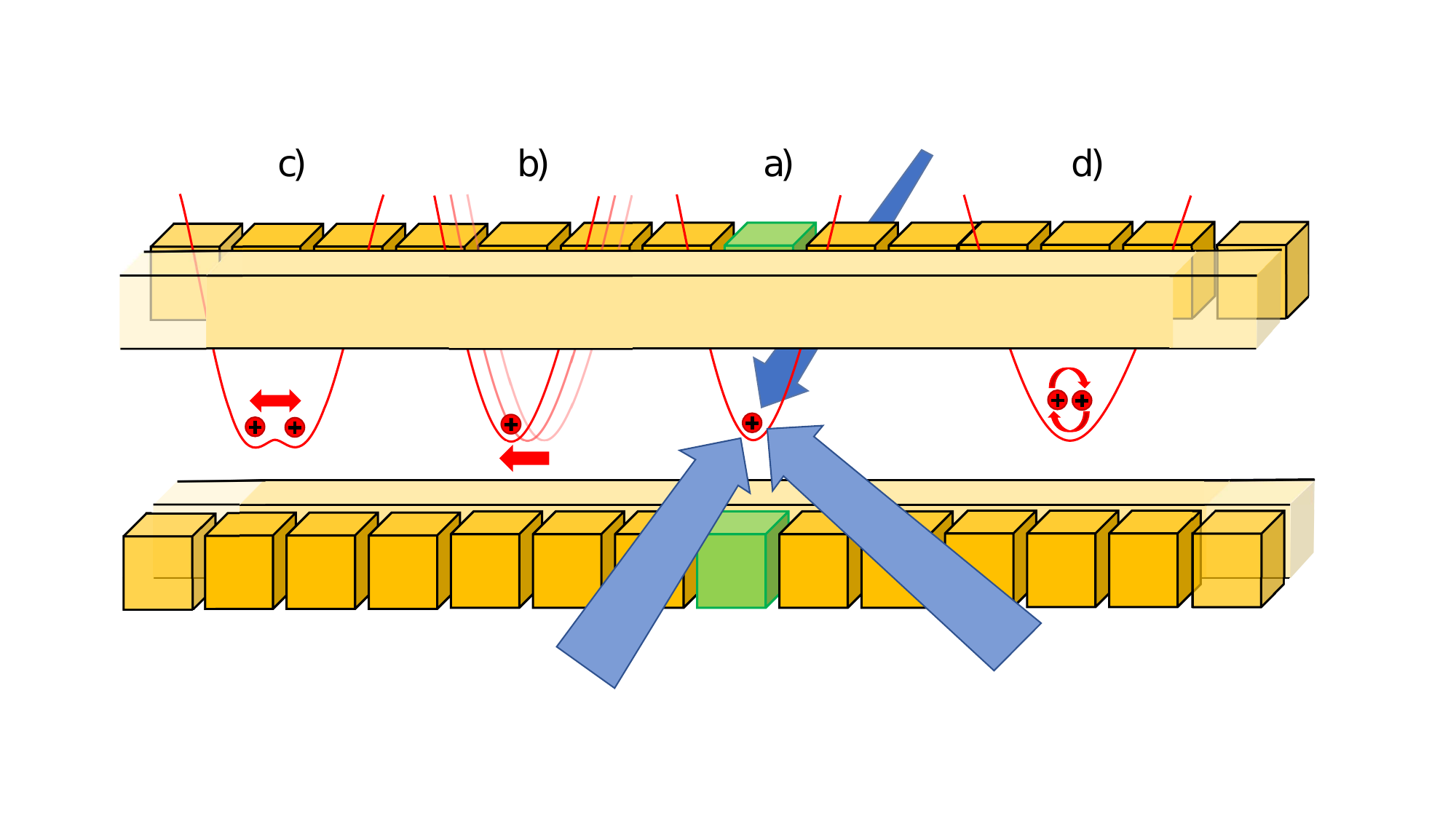}
  	\caption{Architecture of the linear segmented ion trap. Each of the linearly arranged dc segments (orange boxes) can generate a confining electrostatic well (red), capable of storing ion crystals (red spheres). The rf electrodes (elongated yellow boxes) provide confinement in the transverse plane. The laser interaction zone (LIZ) at \textbf{a)} is shown in green, with the blue arrows indicating laser beams. The trap structure continues beyond the left and right ends. Different types of shuttling operations are indicated: \textbf{b)} Linear transport, \textbf{c)} split or merge of ions, and \textbf{d)} crystal rotation.}
    \label{fig:F1}
\end{figure*}

\section{Problem statement}
\label{sect:ProbStatmnt}

The specific shuttling-based ion trap quantum computer architecture considered in the present article is based on a microstructured ion trap consisting of 32 electrode pairs, uniformly arranged along a \textit{trap axis} (Fig. \ref{fig:F1})~\cite{shuttlingIonTrapQuantum}. Each electrode pair can generate a confining potential well. The regime in which no more than two ions are stored simultaneously in each well is considered here, allowing for the best degree of control for gate and shuttling operations, but also leading to maximum shuttling overhead. Hence, the qubit ions are stored individually or in pairs arranged along the trap axis, in segments numbered from~1 to~32. 
As trapped ions within a common well form Coulomb crystals with fixed spatial order, a set of ions stored within one well will be henceforth referred to as a \textit{crystal} (even if the set contains only one ion).
In the present work, one- and two-ion crystals are considered.
Furthermore, the notions \textit{qubit} and \textit{ion} will be used interchangeably from now on. 

Gate operations are carried out using laser beams directed at a fixed site (located at segment~19, 
called the \textit{laser interaction zone} (LIZ). For the architecture considered here, the native gate set consists of local $SU(2)$ qubit rotations on one or two ions and collective $\hat{\sigma}_z\otimes \hat{\sigma}_z$ phase gates on two ions. A local rotation on qubit $j$ is described by the unitary
\begin{equation}
    \hat{R}_j(\theta,\phi)=\exp\left(-i\,\frac{\theta}{2}\,\big(\cos\phi\,\hat{\sigma}^{(j)}_x+\sin\phi\,\hat{\sigma}^{(j)}_y\big)\right) ,
\end{equation}
while simultaneous rotations on qubits $j,k$ are described by $\hat{R}_j(\theta,\phi)\otimes \hat{R}_k(\theta,\phi)$. The two-qubit phase gates are described by
\begin{equation}
    \hat{G}=\exp\left(-i\,\frac{\pi}{4}\,\hat{\sigma}^{(j)}_z\otimes \hat{\sigma}^{(k)}_z\right) .
\end{equation}
These operations are sufficient to generate a universal gate set. Note that different gate sets, for instance employing M{\o}lmer-S{\o}rensen-type (e.g. $\hat{\sigma}_x\otimes\hat{\sigma}_x$) gates are compatible with the scheduling scheme described in the present work, the preceding compilation stage may merely yield a different gate sequence for a given quantum algorithm, leading to different shuttling schedules. Also, our scheme is not necessarily restricted to laser-based gate operations; local operation sites employing near-field microwave radiation as envisioned in \cite{winnie2020} are equally viable.

For the QCCD architecture, shuttling operations are required to implement gate-based quantum circuits. It is assumed that a given circuit to be executed is compiled into the native gate set pertaining to the architecture by a compilation stage done beforehand. The execution of the circuit on the shuttling-based architecture is as follows: First, the required number of ions is distributed within the trap, prepared via cooling and state preparation protocols. A qubit index is initially assigned to each ion, referred to as the \textit{initial ordering}. Then, the actual circuit is executed in repeated steps of i) executing gates on the qubits stored in the LIZ, and ii) reconfiguration of the register via shuttling operations, such that the succeeding qubit(s) on which a gate is to be performed, is stored in the LIZ. Finally, the qubit register is read out.

The shuttling operations required for register reconfiguration consist of manipulating the electrostatic \textit{potential wells} storing the ion crystals. These operations are realized via dynamically changing the voltages applied to the trap electrodes, and fall into three elementary classes:
\begin{compactitem}
\item
\textbf{Linear transport:} A potential well containing a crystal is moved along the trap axis \cite{Walther2012,Bowler2012}. The well is not allowed to cross the position of neighboring wells.
\item
\textbf{Split~/~merge:} Upon crystal split, a single well containing a two-ion crystal is transformed into two distinct wells, partitioning the two ions on these, leading to two one-ion crystals~\cite{Bowler2012,Ruster2014,Kaufmann2014}. Crystal merging is the reverse process.
\item
\textbf{Crystal rotation:} A two-ion crystal is physically rotated, such that the qubit ordering is reversed; this is equivalent to a SWAP quantum logic gate \cite{Kaufmann2017swap}. The rotation-based swap is used as it may lead to different timing overheads and infidelity mechanisms as the gate-based SWAPs. Moreover, rotations can be done outside the LIZ in any segment of the trap, and also at several sites in parallel. This can significantly reduce the overall cost of SWAP operations.
\end{compactitem}

\section{Constraints and costs}
\label{sect:ConstraintsCosts}

The specific hardware details of the platform define physically allowed sequences. Most importantly, the number of trap segments and the number and positions of LIZes determine the layout of the generated shuttling sequences. Furthermore, the trap geometry, fabrication quality and operation parameters limit the number of ions that can be stored within one potential well. In this work, we consider a single LIZ and a maximum number of two stored ions per well.

\begin{figure}
    \centering
 	\includegraphics[width=\columnwidth,trim={0 2.5cm 0 2.5cm},clip]{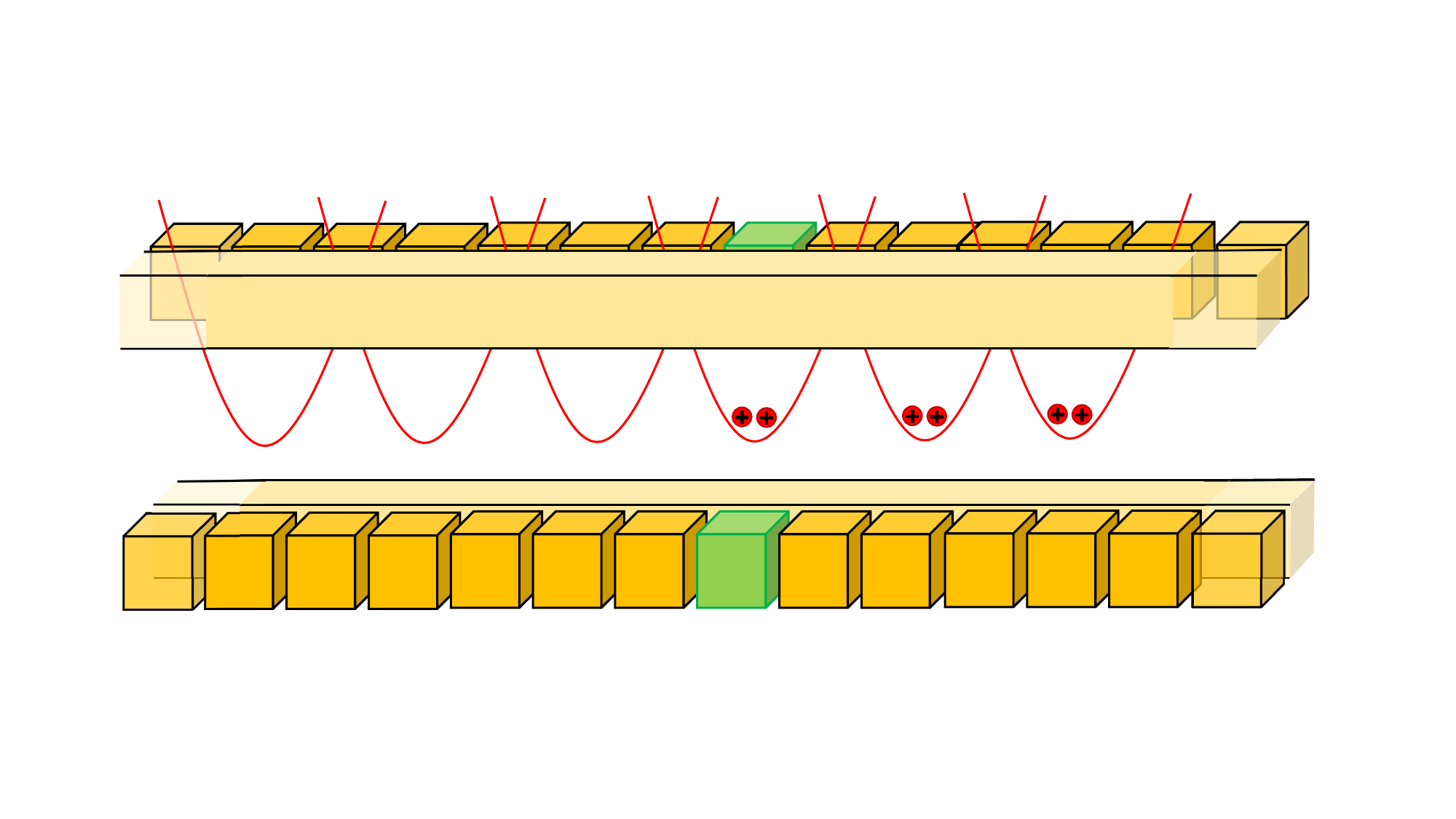}
  	\caption{Empty well positioning before a crystal operation or a gate implementation.}
    \label{fig:F3}
\end{figure}

Further constraints determine which types of shuttling operations can be carried out at which locations. This is of particular relevance, as especially the split~/~merge operations require precise spectroscopic calibration in order to work properly \cite{Ruster2014}, which currently limits their operation to be performed only in the LIZ. A further constraint enforces additional empty potential wells, such that upon sensitive gate or shuttling operations, a balanced electric field configuration at the LIZ is always guaranteed (Fig.~\ref{fig:F3}). The solution approach described in this work is based on a complete list of constraints detailed in Table~\ref{tab:Constraints}.
The constraints determine the maximum total number of qubit ions on which a given architecture can operate. Upon changes of hardware or methodology, the constraints can be adapted, such that solutions matching to different capabilities can be computed straightforwardly.

\begin{table}[t]
  \centering
  \caption{Constraints for the architecture considered herein (Y = yes/allowed, N = no/not allowed).}
  \label{tab:Constraints}
  \small{
  \begin{tabular}{ |c|l|c| } 
 \hline
 \# & Constraint & Setting for \\
    &            & this work \\
 \hline\hline
 1 & Number of trap segments & 32 \\
 \hline 
 2 & Max number ions per crystal & 2  \\
 \hline 
 3 & Number of LIZes & 1 \\
 \hline
 4 & LIZ position (segment) & 19 \\ 
 \hline
 5 & Min space between crystals & 2 \\
   & (segments) &   \\
 \hline
 6 & Empty wells required & Y \\
 \hline
 7 & Split~/~merge outside LIZ & N \\
 \hline 
 8 & Rotation outside LIZ & N \\
 \hline 
 9 & Parallel rotations & N \\
 \hline 
 10 & Max crystal size for rotations & 2 \\
 \hline\hline
\end{tabular}
}
\end{table}

Generating optimal or near-optimal shuttling sequences requires a cost metric to be defined. The time required by the shuttling operations is given by hardware parameters, such as the magnitude of electric fields within the trap, which in turn depend on trap geometry and on the electrical specifications of the arbitrary waveform generators driving the operations \cite{shuttlingIonTrapQuantum}. For current implementations of shuttling-based trapped-ion quantum logic, the shuttling overhead typically exceeds the total time required for quantum logic gates by about one order of magnitude \cite{Pino2021}.
Furthermore, the shuttling operations lead to residual, undesired oscillatory motion of the ions, which can deteriorate the fidelity of subsequent two-qubit gate operations. This suggests that a suitable cost metric can be simply taken as the count of shuttling operations for a given solution. The results presented in this work are based on a cost metric given by the count of split~/~merge operations needed in a shuttling sequence, since these require longer execution times compared to other operations and lead to significantly larger undesired excitation of excess motion \cite{Ruster2014,shuttlingIonTrapQuantum}.

The cost function used in this article does not consider one-qubit gates because the cost of moving crystals in terms of qubit error injection, as considered here, is negligible when compared to the cost of merging and swapping ions. As a consequence, we do not study one-qubit gates in the experiments.
\section{Solution approach}\label{sect:OverallApproach}

For establishing a sequence of commands to realize a given quantum circuit, one first has to assign an ordering to the qubits, to be called the initial ordering. This means assigning a fixed qubit index to an ion stored at a given initial location in the trap.
It is worth noting the importance of the initial ordering, as later alterations to this ordering entail multiple ion permutations in the LIZ, leading to  additional shuttling costs.
The problem of initial qubit placement in the trap segments for a given circuit is comparable to the classical circuit placement problem, which is known to be NP-complete~\cite{NP_Complet_Problem}. This problem grows exponentially with the number of qubits and the size of the architecture, as for $N$ qubits there are $N! /2$ possible different assignments of qubits to ions. The factor $1/2$ any order reversal leads to the same shuttling costs. As the number of possible initial arrangements scale exponentially, a heuristic approach is needed to find an acceptable initial ordering.

Given the complexity of establishing both the initial ion ordering and the optimal sequence of commands for shuttling the ions, we choose to implement a greedy algorithm. All approaches considered proceed in two stages. The first stage consists of providing an initial ordering to start a shuttling sequence. The intent behind the initial ordering is to minimize a priori the number of shuttling operations in the sequence, which is generated in the second stage. In the following, these two stages are addressed sequentially.

\section{Initial ion ordering}
\label{sect:InitialIonOrder}

The purpose of the initial ordering is to associate to each qubit 
in a circuit an ion 
initially loaded and located at a definite position in the segmented trap.
Here, a qubit is an abstraction, whereas an ion is a physical realization of a qubit. In the following, the description will be in terms of qubit positions in the segmented trap, it being understood that a qubit is associated with a physical ion, which itself is located at a definite position in the segmented trap.

Three different initial orderings are explored to investigate the impact of the initial ordering on the cost outcome.
All orderings rely on common data structures (DS) described next.

As mentioned previously in Sect.~\ref{sect:ProbStatmnt}, prior to executing a quantum circuit, the segmented trap is initially loaded with the required number $n$ of qubit ions, with these ions being distributed in the trap.
Software-wise, a first DS models the segmented trap as a one-dimensional array of elements called the \textit{trap model}, the elements of which correspond to the trap segments. The size $S$ of the array is the number of segments of the trap ($S = 32$ here, with the segments indexed from~1 to~32).
The second DS, is a list pertaining to each trap segment, which can be empty or contain one “crystal” DS (see below). 
The LIZ is a special segment in the model at position~19 
in the array of segments.
The third DS is the \textit{crystal} comprising a unique identifier (an integer), an ordered list of ions, with this list containing zero (empty list), one, or two ions (Sect.~\ref{sect:ProbStatmnt}), an integer that specifies the segment in which it is currently positioned.
The fourth DS is the \textit{ion}, which contains a unique identifier (integer) in the range from~1 to~$n$
corresponding to the identifier (integer) of the qubit it encodes, and another integer specifying to which crystal it currently belongs.
When a crystal is split, two new crystals are created with new unique identifiers. Similarly, when two crystals are merged, a new crystal is created with a new identifier.
A fifth DS is the \textit{gate}, which is a structure containing the type of gate and a list of one (1-qubit gate) or two qubits (two-qubit gate) on which the gate operates. 
Finally, a sixth DS is the ordered list of all the gates in the circuit representing the order of the gates as they appear in the circuit. If two gates are simultaneous, an arbitrary order between the two is chosen.
Note that each qubit in the circuit is not initially associated with an ion, and it is one of the purposes of the initial orderings discussed below to establish such an association.

\begin{algorithm}[H]

\caption{Increase pairwise order - IPO}
\label{Al:IPOJD}
\footnotesize{
\begin{algorithmic}[1]
	\STATE{$L$: list of all ions}
	\STATE{$G$: ordered list of all gates}
	\STATE{$V$: list of all crystals created in the first pass (initially \\
	            \hspace{12pt} empty)}
	\STATE{$V'$: final ordered list of all crystals resulting from the second pass (initially empty)}
	\STATE{// First pass, ions are placed in crystals}
  	\FORALL{$g \in G$}
  	\STATE{$I_u,I_v$ := $g$.get\_ions()}
  	\IF{ $I_u,I_v \in L$}
  	\STATE{$V$.append\_crystal($I_u,I_v$)}
  	\STATE{$L$.remove($I_u,I_v$)}
  	\ENDIF
  	\ENDFOR
	\WHILE{$L$.is\_not\_empty()}
		\IF{len($L$) $ \geq  2$}
		\STATE{$V$.append\_crystal($L[0],L[1]$)}
  		\STATE{$L$.remove($L[0],L[1]$)}
  		\ELSE
		\STATE{$V$.append\_crystal($L[0]$)}
  		\STATE{$L$.remove($L[0]$)}
  		\ENDIF
		\ENDWHILE
  	\STATE{// Second pass, final crystal placement}
  	\FORALL{$g \in G$}
  	\STATE{$I_u,I_v$ := $g$.get\_ions()}
  	\IF{$I_u$.do\_not\_share\_crystal($I_v$) and \\
  	    \quad ($I_u$.get\_crystal() $\in V$ or $I_v$.get\_crystal() $\in V$)}
	\IF{$I_u$.get\_crystal() $\in V$ = $I_v$.get\_crystal() $\in V$}
	\STATE{$V'$.append\_crystal($I_u$.get\_crystal(), $I_v$.get\_crystal())}
	\STATE{$V$.remove($I_u$.get\_crystal(),$I_v$.get\_crystal())}
	\ELSIF{$I_u$.get\_crystal()~$\in$~$V$ and \\
	       \quad\quad\quad\, $I_v$.get\_crystal() $\notin$ $V$}
	\STATE{$V'$.place\_closest.($I_v$.get\_crystal(), \\
	   \quad\quad\quad\quad\quad\quad\quad\, $I_u$.get\_crystal())}
	\STATE{$V$.remove($I_v$.get\_crystal())}
	\ELSE
	\STATE{$V'$.place\_closest($I_u$.get\_crystal(), \\
	   \quad\quad\quad\quad\quad\quad\quad\, $I_v$.get\_crystal())}
	\STATE{$V$.remove($I_u$.get\_crystal())}
  	\ENDIF	
  	\ENDIF	
  	\ENDFOR  
\STATE{$V'$.place\_in\_the\_model()}
\end{algorithmic}
} 
\end{algorithm}

The first initial ordering is \textit{order as is} (OAI), in which the qubit indices are simply assigned left-to-right according to the initial ion positions in the segmented trap.
The second approach is \textit{order inputs randomly} (OIR), where the initial ordering is randomly chosen from a uniform distribution. The last approach is the greedy algorithm \textit{increase pairwise order} (IPO). The first two approaches, which are self-explanatory, are primarily used to benchmark the IPO approach, which is described in the form of pseudocode in Algorithm \ref{Al:IPOJD}.

\begin{figure}[t]
    \centering
 	\includegraphics[width=0.50\textwidth]{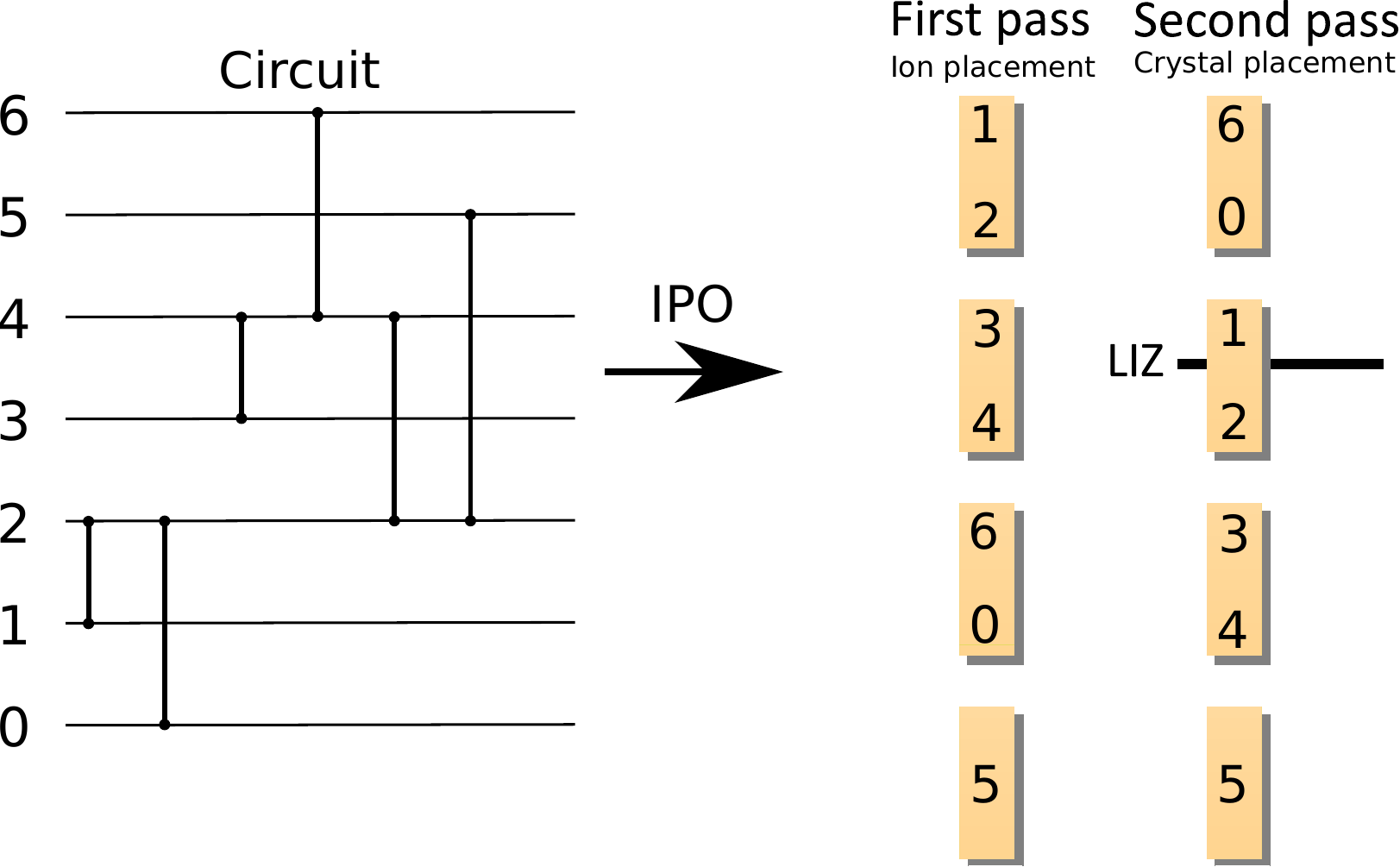}
  	\caption{Example of a conversion from a circuit to an IPO initial ordering. The first pass creates the crystals  containing the ions. The second pass places the crystals with the first ions interacting in the LIZ. 
  	}
    \label{fig:IPO}
	 \vspace{-3mm}
\end{figure}

In algorithms presented in this paper, the symbol "=" represents logical equality between two quantities, and assignment is denoted by the symbol ":=".
Algorithm~\ref{Al:IPOJD} provides a pseudo-code in object-oriented format of the IPO algorithm which proceeds in two passes. The first pass starts from lists $L$, the list of all ions. The ordered list $G$ of all gates is scanned such that ions participating in two-qubit gates are placed in crystals. If one ion is already placed in a crystal, but the other is not, that other ion is placed in a crystal in which an ion is already placed, and if no such crystal exists, a new crystal is created. At the end of the first pass, the list is all ions $L$ is then empty, and all ions are in crystals resulting is the list of crystals $V$. 
In the second pass, the list of all gates $G$ is scanned again in order to reorganize the crystals in $V$ by placing them in an new list $V'$. The list of all gates $G$ is scanned again, identifying gates for which ions are not in the same crystals. These crystals are removed from $V$ and placed adjacent to one another in $V'$. If one of the crystals is already in $V'$, the other crystal is placed in $V'$ at one of the ends that is closest to the already placed crystal. The method place\_closest(c1,c2) is used in these steps; it places the crystal c2 at the end of the $V'$ list closest to the crystal c1.
After all gates have been scanned, if crystals are still remaining in $V$, they are then placed at one end of $V'$.
The method place\_in\_the\_model() that appears at the end of the pseudo-code
maps to each segment a crystal in the list of all crystals $V'$ in such a way that the crystal in the LIZ is the one implementing the first gate. This method does not modify $V'$. The segments where the crystals are placed are determined by the physical constraint of the model. For example, in the ion trap quantum computer considered here, each crystal must be separated from another crystal by one empty segment. The set of physical constraints will be referred to as the constraint set.
Fig.~\ref{fig:IPO} illustrates the IPO algorithm for a very simple case.

\section{Shuttling algorithm}
\label{sect:Shuttling}
\begin{figure}[t]
    \centering
 	\includegraphics[width=0.48\textwidth]{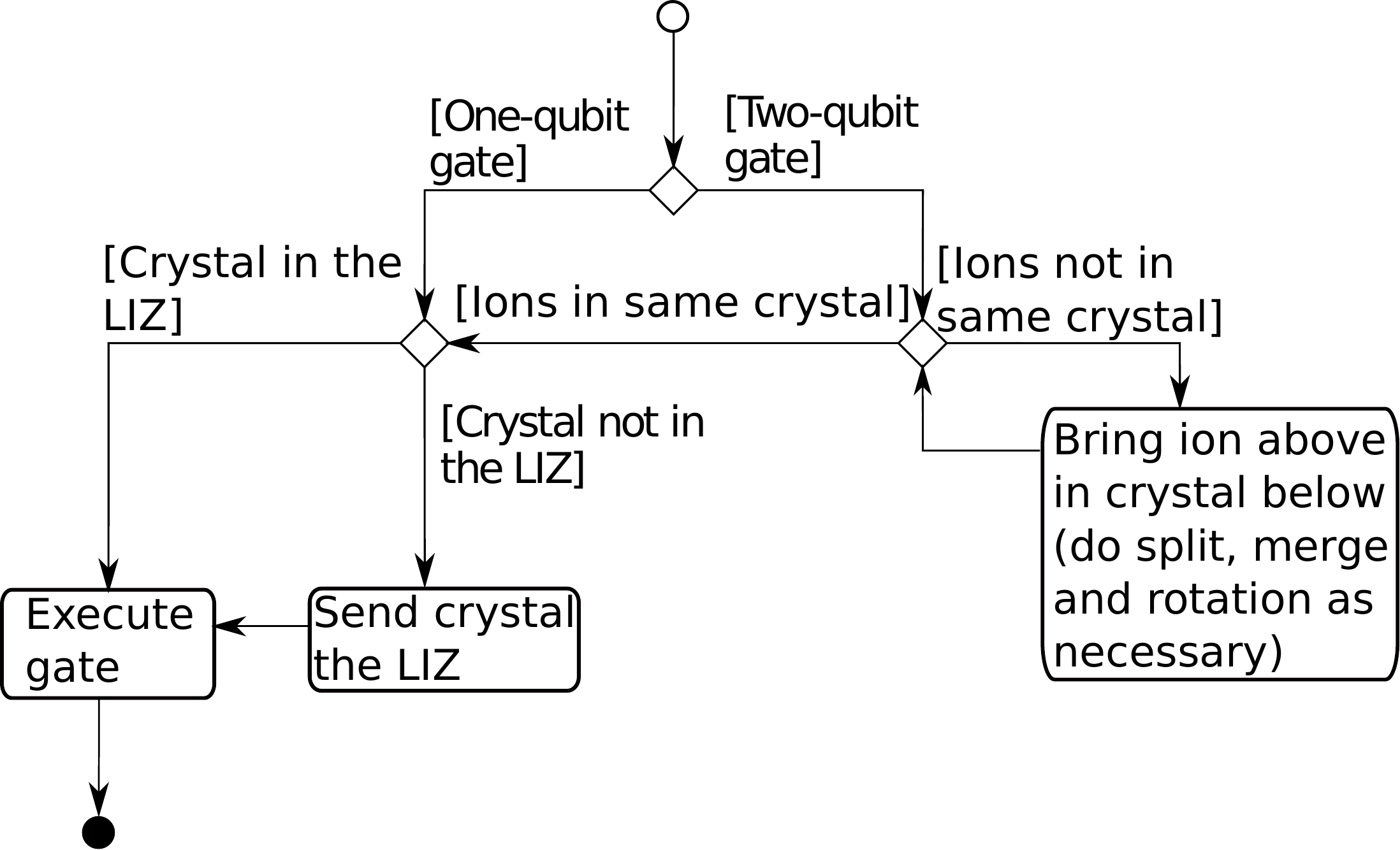}
  	\caption{Algorithm to generate the shuttling sequence for one quantum gate.
             Note: The box "Bring ion above in crystal below" requires splits, merges and rotations to be performed as necessary, see Algorithm~\ref{Al:SHUTTLING})
             }
    \label{fig:algoritmUML}
\end{figure}

If a one-qubit gate is to be executed on a qubit ion not stored in the LIZ, the ion must be moved in the LIZ to execute the gate. In the case of a two-qubit gate, if it involves two ions that are not in adjacent crystals, the ions must be reordered to merge them into the same crystal, and then move that crystal to the LIZ. 
The pseudo-code of the algorithm generating the shuttling schedule is presented in Algorithm~\ref{Al:SHUTTLING}. This algorithm iterates though the gate list $G$, each gate being being process sequentially. For a one-qubit gate, the algorithm is straightforward: first the crystal containing the ion of interest is moved to LIZ if it not stored there already, then the gate is executed.  For a two-qubit gate, if the ions are already in the same crystal, then the same process as for a one-qubit gate is used.
In the following, we denote the shuttling directions with the linear trap as follows: The \textit{top} of the trap is segment $1$, and the \textit{bottom} is the segment at the other end ($32$ for the trap considered herein). Of a given set of ions, the \textit{top} ion is the ion closest to the top segment. An ion is stored \textit{above} another ion if it is closer to the top segment. The same nomenclature applies for the \textit{bottom} / \textit{below} direction.\\
If the ions are in different crystals, then the top ion, is exchanged between crystals until it is in a crystal adjacent to the bottom crystal. Note that if the top ion is at the top within its crystal, the function ion\_permutation (see Algorithm~\ref{Al:Ion permutation}) takes care of moving the crystal to the LIZ and permuting the ions within the crystal by means of a rotation operation, as it is necessary to be in the LIZ for this operation. Similarly, if the bottom ion is at the bottom in its crystal, the crystal is rotated to bring the bottom ion to the top of its crystal. Once the two ions are in adjacent crystals and adjacent to each other, each crystal is split and the ions of interest are merged into a temporary crystal (again this is taken care of in the function ion\_permutation). The gate is then implemented on the temporary crystal. A UML flowchart of the algorithm is depicted in Fig.~\ref{fig:algoritmUML}.

\begin{algorithm}
\caption{Shuttling sequence generation}
\label{Al:SHUTTLING}
\footnotesize{
\begin{algorithmic}[1]
	\STATE{$G$: ordered list of all gates}
	\STATE{DoGate: Boolean to implement or not a gate}
\FORALL{$g \in G$}
	\IF{$g$.is\_one\_qubit\_gate()}
		\STATE{$C$ := $g$.get\_ion().get\_crystal()}
		\IF{$C$.is\_in\_LIZ() = False}
		\STATE{$C$.send\_to\_segment(LIZ)}
		\ENDIF
        \STATE{g.LIZ\_implement\_gate()}
	\ENDIF
	
	\IF{$g$.is\_two\_qubit\_gate()}
		\STATE{$I_1$,$I_2$ = $g$.get\_ions()}
		\IF{$I_1$.get\_crystal()~$\neq$~$I_2$.get\_crystal()}
		\STATE{DoGate:= False}
    		\WHILE{$I_1$.get\_crystal\_below()~$\neq$~$I_2$.get\_crystal()}
        \STATE{$C$~:=~$I_1$.get\_crystal().get\_crystal\_right()
    		\STATE{ion\_permutation($I_1$,$C$.get\_left\_ion(),DoGate)}}
    		\ENDWHILE
    		\STATE{DoGate:= True}
    		\STATE{ion\_permutation($I_1$,$I_2$,DoGate)}
    		\ELSE
    			\STATE{$C$ = $I_1$.get\_crystal()}
    			\IF{$C$.is\_in\_LIZ() = False}
    			\STATE{$C$.send\_to\_segment(LIZ)}
    			\ENDIF
        		\STATE{g.LIZ\_implement\_gate()}
    		\ENDIF
	\ENDIF
\ENDFOR
\end{algorithmic}
} 
\end{algorithm}

The send\_to\_segment(seg) method appearing in the shuttling algorithm (Algorithm~\ref{Al:SHUTTLING}) is a specific method that sends a crystal to segment \textit{seg}. The method is detailed in Algorithm~\ref{Al:Move Crystal}.If the crystal meets another crystal on its way to the LIZ, then this other crystal is sent recursively via the same algorithm to the segment above or below  the segment \textit{seg} (the direction depends on the direction of motion of the former crystal). The method move\_to\_segment(seg) appearing in Algorithm~\ref{Al:Move Crystal} moves the crystal along the segments to bring it to a target segment \textit{seg} after the way has been cleared between the crystal and the target segment.

\begin{algorithm}
\caption{Algorithm send\_to\_segment.}
\label{Al:Move Crystal}
\footnotesize{
\begin{algorithmic}[1]
\STATE{// It is assumed that the crystal is positioned at a segment above the targeted segment; the algorithm is easily changed if the crystal is below the targeted segment (different direction).}
\STATE{$S_{targ}$: targeted segment}
\STATE{$C_{mov}$: moving crystal}
\WHILE{$C_{mov}$.get\_segment() $\neq$ $S_{targ}$}
    \IF{$C_{mov}$.get\_segment\_left().is\_occupied()}
        \STATE{$C$ =: crystal.get\_segment\_left().get\_crystal()}
        \STATE{// Recursive call of the algorithm}
        \STATE{$C$.send\_to\_segment(segment.segment\_left())}
    \ENDIF
    
    \STATE{$C_{mov}$.move\_to\_segment($S_{targ}$)}
\ENDWHILE
\end{algorithmic}
}
\end{algorithm}

The ion permutation function appearing in Algorithm~\ref{Al:Ion permutation} exchanges two ions between two crystals. The algorithm creates a temporary crystal on which the two-qubit gate is to be executed. This function is the main contributor to the cost in implementing a circuit because it involves the split and merge commands. The steps of the algorithm are considered to be self explanatory and no further description of this algorithm will be provided here.

\begin{algorithm}
\caption{Algorithm ion\_permutation.}
\label{Al:Ion permutation}
\footnotesize{
\begin{algorithmic}[1]
\STATE{$C_1$ : Crystal 1}
\STATE{$C_4$ : Crystal 4}
\STATE{DoGate: Boolean to implement a gate or not (true if a gate is to be executed on the ions that are permuted}
\STATE{// Crystal $C_1$ is considered above crystal $C_4$, the objective is to exchange ion $I_1$ in $C_1$ with ion $I_2$ in $C_4$. Note: Crystals $C_2$ and $C_3$ are created during the process.}
\IF{$I_1$.is\_left()}
    \STATE{$C_1$.rotate()}
\ENDIF
\IF{$I_2$.is\_right()}
    \STATE{$C_4$.rotate()}
\ENDIF

\STATE{// Reminder: The result of a split gives two crystals of one ion each, ion $I_1$ being in crystal $C_2$.}
\IF{$C_1$.get\_number\_ions() = 2}
\STATE{$C_1$.send\_to\_segment(LIZ)}
\STATE{$C_1$,$C_2$ := LIZ.split()}
\ELSE
\STATE{$C_2$ := $C_1$}
\ENDIF
\IF{$C_4$.get\_number\_ions() = 2}
\STATE{$C_4$.send\_to\_segment(LIZ)}
\STATE{$C_3$,$C_4$: = LIZ.split()}
\ELSE
\STATE{$C_3$ := $C_4$}
\ENDIF

\STATE{$C_2$.send\_to\_segment(LIZ-1)}
\STATE{$C_3$.send\_to\_segment(LIZ+1)}

\STATE{$C_2$ := LIZ.merge()}
\STATE{$C_2$.rotate()}
\IF{DoGate = True}
\STATE{LIZ.execute\_gate()}
\ENDIF
\STATE{$C_2$,$C_3$ := LIZ.split()}

\IF{$C_1$.get\_number\_ions() = 2}
\STATE{$C_1$.send\_to\_segment(LIZ-1)}
\STATE{$C_2$.send\_to\_segment(LIZ+1)}
\STATE{$C_1$ := LIZ.merge()}
\ENDIF
\IF{$C_4$.get\_number\_ions() = 2}
\STATE{$C_3$.send\_to\_segment(LIZ-1)}
\STATE{$C_4$.send\_to\_segment(LIZ+1)}
\STATE{$C_4$ := LIZ.merge()}
\ENDIF
\end{algorithmic}
}
\end{algorithm}

\begin{figure*}[t]
    \centering
    \vspace{-1cm}
    \subfloat[]{
    \label{FigureInputCompiler}
    \includegraphics[width=0.15\textwidth]{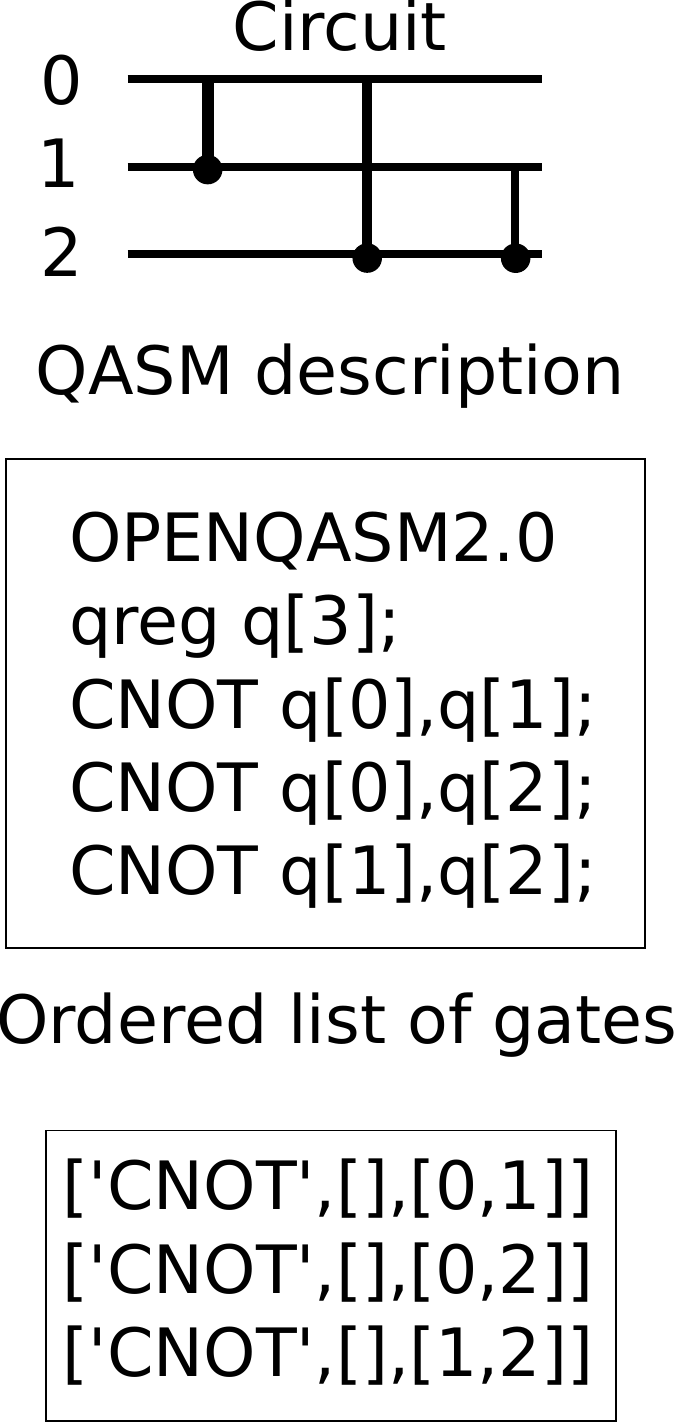}}
    \quad
    \subfloat[]{
    \label{FigureTabularCompiler}
    \includegraphics[width=0.3\textwidth]{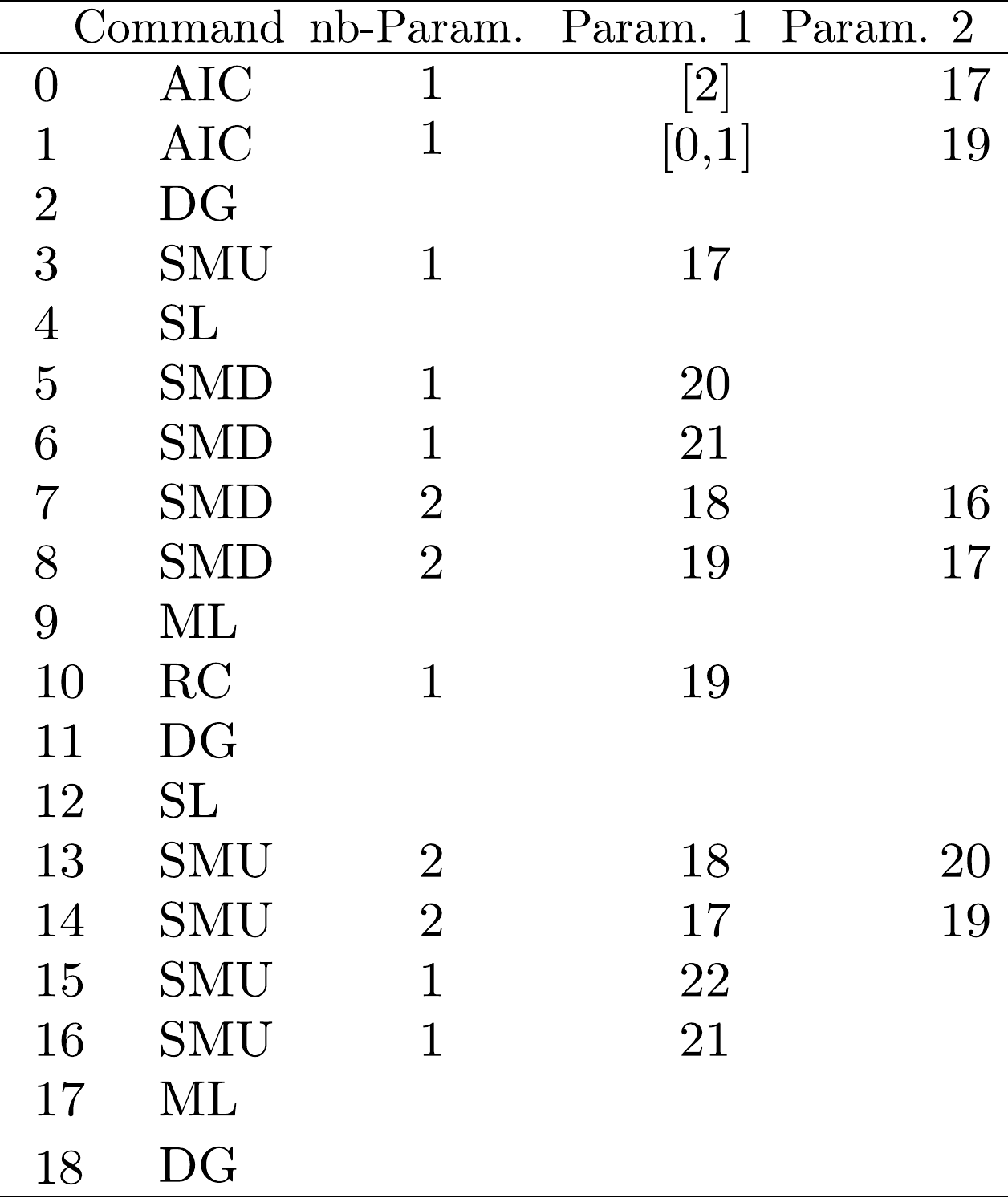}}
    \quad
    \subfloat[]{  \label{FigureGraphicalCompiler} 
    \includegraphics[width=0.38\textwidth]{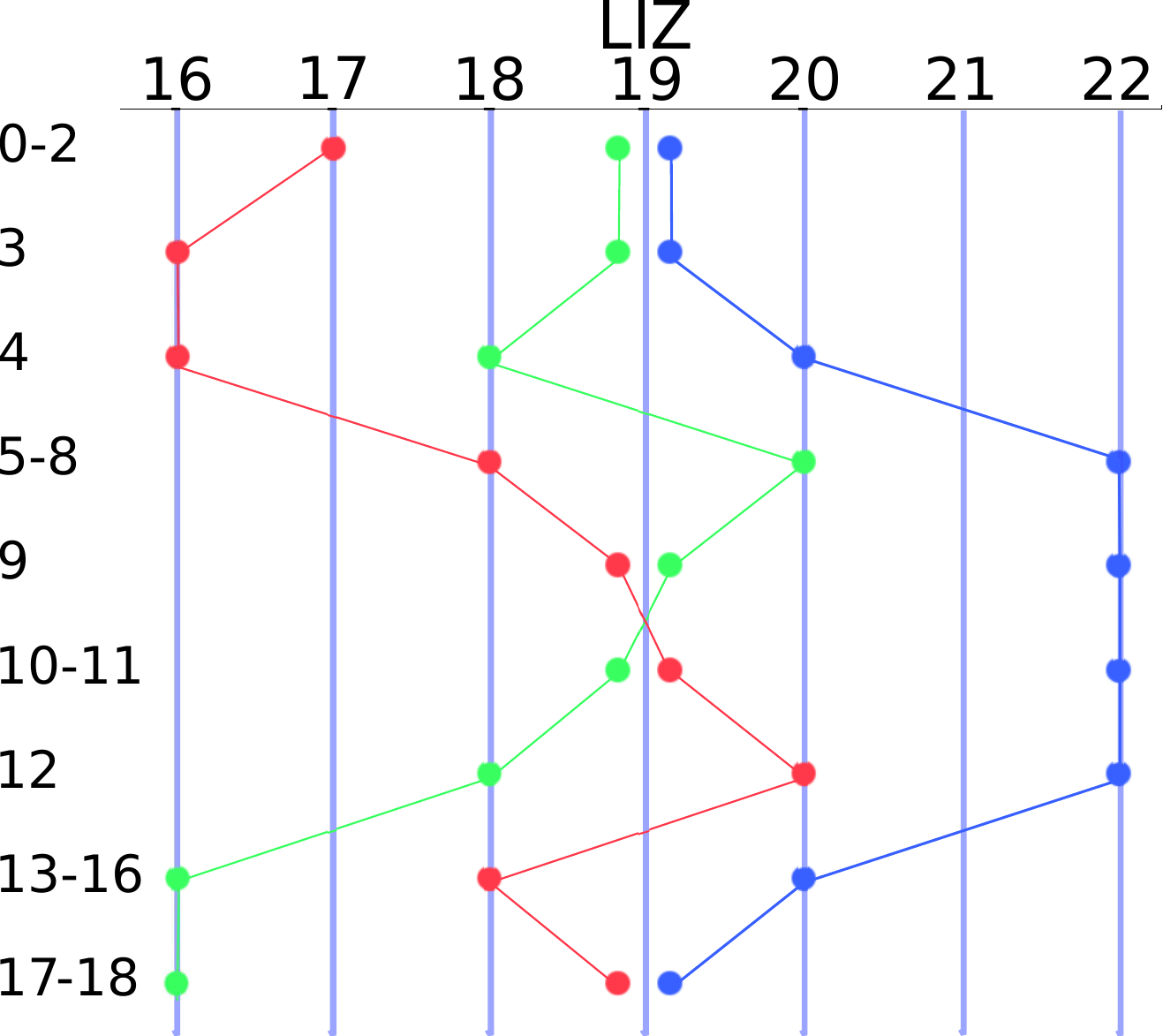}}
    \caption{
    Shuttling sequence generation example. 
    \textbf{(a)}~ Example circuit, its QASM description, and the ordered list of gates generated by the parser.
    \textbf{(b)}~Table of instructions for a shuttling sequence as generated by the compiler (see Table \ref{tab:ShuttlingCmds}). The first column indicates the sequential number of the command identified as indicated in the second column. The third column indicates the number of parameters that the command receives. 
    \textbf{(c)}~Visualization of the generated shuttling sequence (temporal order is up to down): ion~0 in green, ion~1 in blue, and ion~2 in red. The number identifiers of the commands applied (left column in~(b)) appear at the left of the graph. Some command identifiers are grouped (e.g. 5-8) either because they are displayed at once in the graph (sequence of SMDs as in 5-8), or because they are not visible in the shuttling, such as performing a gate (DG).}
    \label{FigureCompilerCircuit}
\end{figure*}

\begin{table*}[t]
  \centering
  \caption{List of shuttling commands.}
  \label{tab:ShuttlingCmds}
  \small{
  \begin{tabular}{|l|l|l|}
    \hline
    Command & Parameters & Description \\
    \hline\hline
    START & - & Start the shuttling sequence \\
    \hline
    AIC & id-Ion id-Seg & \makecell[l]{Add a crystal with ion id-Ion at the id-Seg
                                      segment in the \\
                                      model. This command can only be used for the
                                      initial ion \\
                                      placement, before shuttling starts.} \\
    \hline
    AEC & id-Seg & Add an empty potential well at the id-Seg segment. \\
    \hline
    REC & id-Seg & Remove an empty potential well at the id-Seg segment. \\
    \hline
    SMU & num-Seg (id-Seg)+ & \makecell[l]{Move up (\textit{i.e.} towards the smallest
                                           segment identifier) the \\
                                           crystals positioned at all listed  
                                           id-Seg. There are num-Seg \\
                                           listed segments.} \\
    \hline
    SMD & num-Seg (id-Seg)+ & \makecell[l]{Move down (\textit{i.e.} towards the largest
                                           segment identifier) the \\
                                           crystals positioned at all listed  
                                           id-Seg. There are num-Seg \\
                                           listed segments.} \\
    \hline
    RC & id-Seg & Rotate the crystal at id-Seg. \\
    \hline  
    ML & - & \makecell[l]{Merge in the LIZ the two crystals in segments \\
                          neighboring the LIZ on each side.} \\
    \hline  
    red SL & - & \makecell[l]{Split the crystal in the LIZ into two one-ion
                         crystals, \\
                         one above and one below the LIZ.} \\
    \hline  
    DG & - & \makecell[l]{Execute gate(s): this is a macro placeholder \\
                          for a sequence of gate instructions.} \\
    \hline  
    \end{tabular}
    } 
\end{table*}

\section{Implementation details}
\label{sect:Implement}

To fully automate the processes of finding the initial ordering and generating a shuttling sequence, a compiler has been implemented using Python 3.7 and an ANTLR4-generated parser~\cite{ANTL4}. The user interface was created with PyQt~\cite{PYQT}. 
The input to the compiler is an OpenQASM~2.0 text file~\cite{cross2017open,OPENQASM2.0}. The parser is used to read and validate the OpenQASM quantum circuit description, producing an abstract syntax tree (AST) as its internal structure. This serves as the basis for generating an 
initial ion ordering, along with the shuttling and gate sequences.
The output of the compiler is a description of the initial ion ordering in the segmented trap (\textit{i.e.} the correspondence between qubits and ions), and the sequence of ion shuttling commands to implement the quantum circuit. Fig.~\ref{FigureCompilerCircuit} shows an example, including a simple circuit along with its OpenQASM representation, the ordered list of gates parsed from this description, the generated code, the shuttling sequence, and a visual representation thereof.
Note that in ion trap quantum computers CNOT gates are not native, and so are realized with gates that are native.  
Realizing the CNOT gate with native gates does not require any additional shuttling, only additional local gate rotations which do not impact the shuttling overhead cost\cite{Kreppel2022}.
As CNOT gates are ubiquitous in the literature, they are used as such in the example depicted in Fig.~\ref{FigureCompilerCircuit}.

The data structure for modelling the state of the ion trap for the purpose of generating shuttling sequences is the trap model described in Sect.~\ref{sect:InitialIonOrder}. The state of the trap at any given time consists of the positions of the ions within the trap.
Algorithm~\ref{Al:SHUTTLING} performs modifications to the state of the trap and each of which corresponds to a shuttling command. Each shuttling command is added sequentially to a list corresponding to the shuttling sequence.
The feasibility of each command is verified for compatibility against the constraint set, an exception being raised when that check fails, with an error message that prompts the user to take an appropriate corrective action (this may happen for example if the number of ions is higher than the trap capacity).
Further downstream in the control software stack, each command in the shuttling sequence is translated into hardware commands. This translation is highly hardware specific and not described here.

A unique identifier is assigned to each command after it has been validated.
As some commands (such as rotate or move) can be carried out in parallel on multiple segments, they are parametrized by a list of segment identifiers. The shuttling sequence commands are listed in Table~\ref{tab:ShuttlingCmds}. 

\section{Results}\label{sect:Results}
\begin{figure}[t]
    \centering
 	\includegraphics[width=0.5\textwidth]{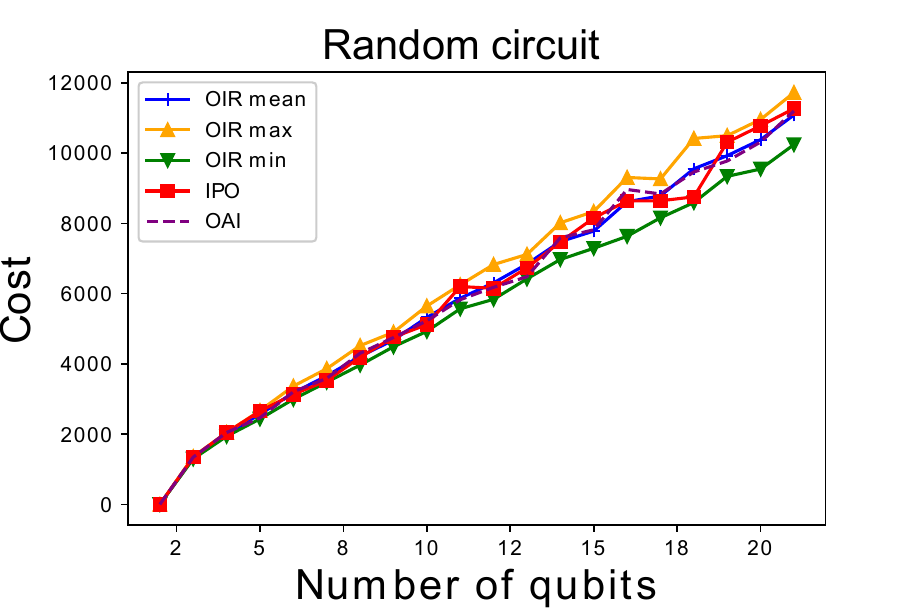}
  	\caption{Cost for random circuits (1000 gates each) versus the number of qubits. Results are presented for: order as is (OAI), order inputs randomly (OIR), and increase pairwise order (IPO).}
    \label{fig:random}
\end{figure}

The different algorithms for generating shuttling sequences were first tested on random quantum circuits for different qubit numbers and a fixed depth of 1000 gates. For the sake of comparison and standardization of the tests, only circuits composed of two-qubit gates have been considered. Random circuits are generated by successively picking two random qubits to undergo a gate from a uniform distribution, this being repeated for 1000 gates. For benchmarking the OIR approach for the initial ordering, 1000 random initial crystal orderings are considered for each random circuit.  A shuttling sequence is generated using the compiler for every random circuit, for each of the 1000 initial ion orders.
The following initial orders have been conside\-red : OAI, OIR, and IPO (Section~\ref{sect:InitialIonOrder}).
OAI and OIR serve as benchmarks against which IPO is compared. Fig.~\ref{fig:random} shows the cost expressed as the number of merge and split commands. As the OIR approach is based on randomization, the graph shows the average, maximum and minimum cost of the solutions as a function of the number of qubits in the circuit. As there are $n!$ possible input orders, the 1000 input orders considered for OIR represent only a very small sample of these possibilities. Therefore, reported minimum and maximum costs are taken from that small sample and are not absolute.
\begin{figure}[t]
    \centering
 	\includegraphics[width=0.50\textwidth]{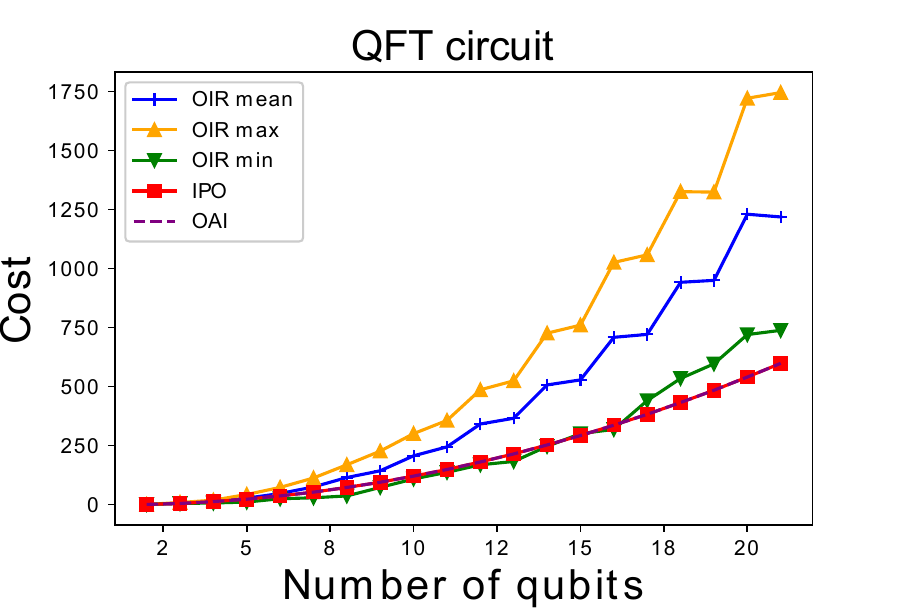}
  	\caption{Cost versus number of qubits for a QFT circuit.}
    \label{fig:QFT}
\end{figure}
\begin{figure}[t]
    \centering
\begin{tikzpicture}
\begin{yquant}
qubit {$ctrl_\idx$} ctrl[5];
qubit {$anc_\idx$} anc[4];
qubit {$target$} target[1];
cnot anc[0] | ctrl[0], ctrl[1];
cnot anc[1] | ctrl[2], anc[0];
cnot anc[2] | ctrl[3], anc[1];
cnot anc[3] | ctrl[4], anc[2];
cnot target[0] | anc[3];
cnot anc[3] | ctrl[4], anc[2];
cnot anc[2] | ctrl[3], anc[1];
cnot anc[1] | ctrl[2], anc[0];
cnot anc[0] | ctrl[0], ctrl[1];
\end{yquant}
\end{tikzpicture}
  	\caption{Implementation of the $N$-Toffoli gate~\cite{Nielsen}, $N$ being the total number of qubits involved in the circuit, $ctrl_i$ being the control qubits, $anc_i$ being ancillary qubits and $target$ being the target qubit.}
    \label{fig:F8}
\end{figure}
\begin{figure}[t]
    \centering
 	\includegraphics[width=0.50\textwidth]{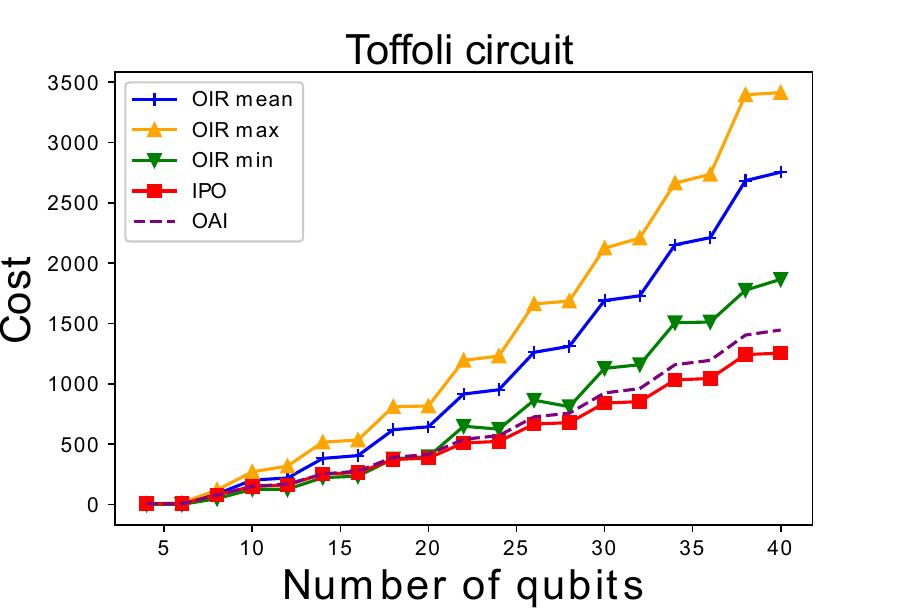}
  	\caption{Cost versus the number of qubits for a Toffoli circuit.}
    \label{fig:F9}
\end{figure}

It is apparent from Fig.~\ref{fig:random} that OIR yields the best results, while the IPO approach does not produce the optimum solution. This is expected, since random circuits do not contain any structure, leading to any systematic optimization approach being of rather limited benefit.

The shuttling overheads for realizing quantum Fourier transform (QFT,~\cite{coppersmith2002approximate}) circuits for varying qubit number are shown in Fig.~\ref{fig:QFT}. Understanding how to best implement this circuit with segmented ion traps can shed light on better generalized heuristics. 
The results indicate that the IPO algorithm produces better solutions as compared to the OIR and OAI approaches. 

Next, we consider a multi-qubit Toffoli gate, depicted in Fig.~\ref{fig:F8}. , which is an important quantum logic primitive. The results are depicted in Fig.~\ref{fig:F9} and shows that the IPO approach outperforms OIR, however IPO yields only slightly better results as OAI, the reason being that the standard circuit representation of the multi-qubit Toffoli gate already arranges the qubit in a natural ordering offering low costs of reconfiguration.

Because the shuttling algorithm uses a greedy heuristic, which does not resort to looking ahead as in classical modern compilers, the time cost of the algorithm for calculating the shuttling sequence is directly proportional to the number of commands output by the algorithm to the machine.

To summarize, we considered well-known circuits which are maximally ordered (QFT and Toffoli), along with random circuits, which represent random unitary transformations and are completely  disordered. In this sense, extreme cases have been covered.
 
\section{Discussion and proof}
\label{sect:DiscussProof}
\begin{figure}[t]
    \centering
 	\includegraphics[width=0.50\textwidth]{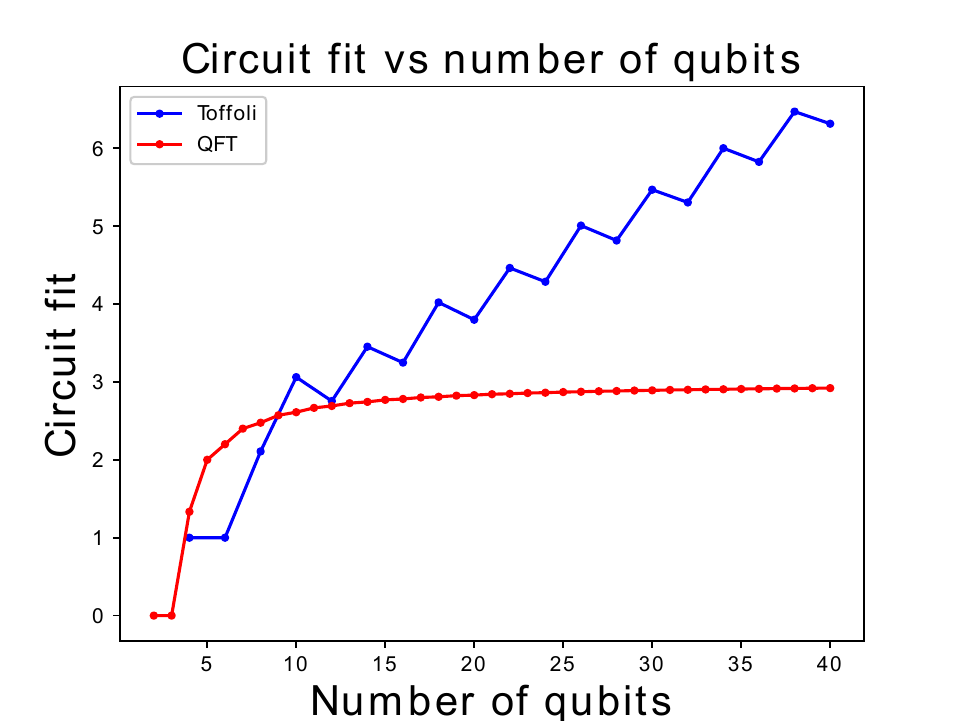}
  	\caption{Circuit fit versus the number of qubits for QFT and Toffoli circuit.}
    \label{fig:CircuitFit}
\end{figure}

Our results show that for the QFT and generalized Toffoli circuits, the IPO approach does not significantly outperform the OAI approach, where the qubits are assigned to physical ions in the order as prescribed by the canonical circuit definitions. However, this requires careful interpretation: First, such circuits typically occur as parts of larger circuits, where the input qubits are \textit{not} necessarily ordered according to the textbook definitions. Here, a systematic ordering as computed by the compilation layer can be used to condition the register in way which reduces the shuttling overhead for conducting the particular subcircuit. Second, for more complex circuits as compared to the generalized Toffoli or QFT examples, a canonical qubit arrangement will most likely not be obvious, such that assignment approaches such as IPO are ultimately required. 
To better understand the impact of the standard descriptions of the QFT and Toffoli circuits, a closer analysis will now be made of their intrinsic structures and the features which make their usual schematic representations induce good initial ordering candidates of the ions in the segmented trap. For this purpose, a metric called the \textit{circuit fit} and denoted by $C$ is introduced and defined as:
\[
  C = \frac{\text{Cost}}{\text{Number~of~gates}}.
\]
The circuit fit is a measure of the mean cost per gate, where the cost is associated with the number of merge/split commands. This metric quantifies how well a shuttling sequence generated by the algorithm is adapted to a given circuit.
Figs.~\ref{fig:CircuitFit} show the results when using this metric on the QFT and Toffoli circuits.

The circuit fit for the QFT asymptotically reaches the value $C=3$ when increasing the number of qubits (ions). In contrast, for the Toffoli circuit, the circuit fit grows with the number of qubits. The reason for this is that the ions are disorganized with respect to the circuit and are never positioned favorably between consecutive gate operations. 
Therefore, the asymptotic behavior of the circuit fit can be seen as a measure of the disorganization of a linearly arranged qubit register with respect to the circuit. Thus, the existence of asymptotes for some types of circuits indicates that circuit structure optimization at compilation stage, as described for example in~\cite{6560634}, may reduce the total shuttling cost.\\

\begin{figure}[t]
    \centering
\begin{tikzpicture}
\begin{yquant}
qubit {$I_\idx$} I[4];
cnot I[1] | I[0];
cnot I[2] | I[0];
cnot I[3] | I[0];
\end{yquant}
\end{tikzpicture}
  	\caption{Example of a circuit with a structure equivalent to a subset of the QFT algorithm. The ion $I_0$ is involved with every gate of the circuit.}
    \label{fig:CirEqQFT}
\end{figure}
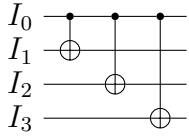

The following will demonstrate the existence of an asymptotic limit for the circuit fit $C$ as a function of the number of ions for the QFT circuit. More precisely, it will first be shown that there is an optimal ordering for shuttling ions in a linear segmented trap for registers of arbitrary size in the case of a QFT circuit. Although our approach is currently limited to storing at most two ions per segment, a more general model will first be considered, where ion traps can be as large as can be. In particular, a system consisting of only two segments will first be considered, with each segment containing half the total number of ions. The result is then extended to prove the existence of a limit $L$ for the circuit fit $C$ for a circuit with a structure equivalent to a subset of a QFT circuit, where one qubit interacts with all the other qubits, as shown in Fig.~\ref{fig:CirEqQFT}. This demonstration is concluded by extending this limit to a larger more general set of circuits.
This demonstration applies to circuits where the gates can be organized in a structure similar to that for QFT operations.

First, consider a circuit $Q$ implementing QFT calculations, where all qubits eventually interact with all other qubits through associated quantum gates. The circuit $Q(I,G)$ is defined as the association of two sets: the set $I$ of ions
\[
  \{I_1, I_2,\ldots I_n\},
\]
and the set $G$ of quantum gates
\[
  \{G_{12},\ldots, G_{1n}, G_{23},\ldots, G_{2n},\ldots,G_{ij},\ldots, G_{(n-1)n}\},
\]
where for every pair of ions $(I_i, I_j)$ where $i\neq j$ in set $I$ there is a gate $G_{ij}$ in $G$.
Further, a partition 
\[
H = \{H_1,H_2,\ldots,H_{(n-1)}\}
\]
of $Q$ is defined, where every subset $H_i$ contains ions 
\[
\{I_i, I_{i+1},\ldots,I_n\}
\] 
and gates 
\[
  \{G_{i(i+1)}, G_{i(i+2)}, \ldots,G_{in}\}.
\] 
Therefore, every subset $H_i$ contains a single ion $I_i$ which interacts with all the other ions of $H_i$, (\textit{i.e.} $\{I_{i+1},\ldots,I_n\}$) through quantum gates
\[
  \{G_{i(i+1)},\ldots,G_{in}\}.
\]
Ion $I_i$ will be labeled as $I_C$ to indicate that it is common to all the gates in $H_i$.
This partitioning structure allows the sequencing of ion interactions in an orderly way such that ion $I_C$ is moved gradually from the first gate $G_{i(i+1)}$ to the last gate $G_{in}$ in $H_i$.

One such subset $H_i$ of $Q$ will now be considered. Without loss of generality, $H_i$ is assumed to contain $m$ ions, relabeled $\{I_1, I_2, ... I_m\}$ with corresponding relabeled gates $\{G_{12}, G_{13},\ldots,G_{1m}\}$; ion $I_1$ will be labeled as $I_C$, the common ion in subset $H_i$.

Ions $\{I_C,I_2,\ldots,I_m\}$ are first assumed to be split between two crystals of size $m/2$:
\begin{equation}
\{I_C, I_2,\ldots, I_{m/2}\}\{I_{m/2+1}, I_{m/2+2},\ldots, I_m\}.
\label{eqSys}
\end{equation}
As ions $I_C$ to $I_{m/2}$ appear in the same crystal, gates $G_{C2}, G_{C3},\ldots, G_{C(m/2)}$ can all be effected without any split/merge operation.
The associated cost in terms of split/merge's is therefore zero. 

Next, ions $I_C$ and $I_{m/2+1}$ are regrouped into a 2-ion temporary crystal,
\begin{equation}
\{I_C,I_{m/2+1}\}.
\end{equation}
where the full linear trap now contains crystals:
\begin{equation}
\{I_{2},...,I_{m/2}\},\{I_C,I_{m/2+1}\},\{I_{m/2+2},...,I_{m}\}.
\label{eqSys2}
\end{equation}
The transition from the initial state (Eq.~\ref{eqSys}) to the intermediate state (Eqs.~\ref{eqSys2}) requires two splits (ion $I_C$ extracted from the first crystal and ion $I_{m/2+1}$ extracted from the second crystal) and one merge (creation of the temporary crystal with ions $I_C$ and $I_{m/2+1}$), that is, 3 split/merge operations.

The gate involving $I_C$ and $I_{m/2+1}$ is then performed and followed by split/merge operations which produce the following configuration:
\begin{equation}
\{I_{2}\ldots,I_{m/2},I_{m/2+1}\},\{I_C,I_{m/2+2},\ldots,I_{m}\}.
\label{eqSys3}
\end{equation}
Similarly to the previous transition from the initial to the intermediate state (Eqs.~\ref{eqSys} and \ref{eqSys2}), transitioning from the intermediate to the final state (Eqs.~\ref{eqSys2} and \ref{eqSys3}) also requires 3 split/merge operations. As a result, the transition from the initial state to the final state (Eqs.~\ref{eqSys} and \ref{eqSys3}) requires a total of 6 split/merge operations.

In the final state (Eq.~\ref{eqSys3}), the operations on the remaining gates of the circuit can then be performed again with zero cost, as no further split\-/merge operations are necessary for the gates involving ion $I_C$ and the remaining ions, since all are then included in the second crystal.

As a result, for a circuit where one ion interacts in sequence with all others, as is the case in subset $H_i$ of a QFT circuit,  the analysis above represents the minimum number of split/merge operations necessary to go over the entire circuit.
In this case, the total split/merge cost is constant (6 split/merge operations). As the number of quantum gates is one less than the number of ions, the circuit fit $C$ is therefore:
\begin{equation}
  C = \frac{6}{m-1}.
\end{equation}
This result can easily be extended to situations where crystals contain $K$ ions instead of $m/2$. In this case, the number of crystals will be $m/K$, and there will be $\frac{m}{K}-1$ intermediate states, as ion $I_C$ is gradually moved from the first crystal to the last crystal, passing through all intermediate crystals on the way. The creation of every intermediate state requires a constant cost of 6, as established above. Therefore, the circuit fit now becomes:
\begin{equation}
  C = \frac{6(\frac{m}{K}-1)}{m-1}.
\end{equation}
As the number of ions increases to infinity, on gets:
\begin{equation}
L = \lim_{m\to\infty} C = \frac{6}{K}.
\end{equation}
For the architecture considered, $K = 2$ and therefore $L = 3$, a value confirmed by simulations (Fig.~\ref{fig:CircuitFit}).
It can also be observed that $L$ can be smaller than 3 only when the number of gates is greater than $(m-1)$ or else when $K$, the number of ions per crystal, is greater than 2.

The analysis above demonstrates that for every subset $H_i$ of $Q$ (a QFT circuit), the last ion configuration conserves the order of all ions other than $I_C$. Therefore, those ions, and all further subsets $H_{i+1}, H_{i+2},\ldots,H_{n-1}$, are still correctly ordered for the following operations. As a result, when using the OAI algorithm for the initial ordering, the circuit fit will approach $L = \frac{6}{K}$ as the number of qubits increases in the QFT.

As the OAI produces the optimal ion ordering for QFT circuits, it is interesting to analyze under which conditions this result can be extended to other types of circuits.  In particular, the case where it is possible to partition a circuit $Q'$ into subsets $H'_i$ where there can be more than one common ion. $H'_i(I'_i, G'_i)$, or more simply $H'(I',G')$ is in this case, defined as follows:
\begin{equation}
I' = \{I_{C_1}, I_{C_2},\ldots,I_{C_w}, I_x, I_{x+1},\ldots,I_m\},
\end{equation}
\begin{equation}
G' = G_{uv},
\end{equation}
 where $u \in \{I_{C_1},\ldots,I_{C_w}\}$ and $v \in \{I_x, I_{x+1}, I_m\}$.
$I'$ is then further partitioned into ion subsets $B_u$ where the set of all common ions is defined as $B_C = B_1 = \{I_{C1}, I_{C_2},\ldots,I_{C_w}\}$, and where other subsets $B_u$ are defined as groups of ions involved only in gates with common ions and themselves:
\begin{equation}
B_u = \{I_a, I_b, I_c,\ldots\},
\end{equation}
such that $G'$ only contains gates $G_{uv}$, with $I_u$ and $I_v \in B_C \cup B_u.$
In that case, if
$|B_C| + |B_u| < K$ for all $B_u \in H'$, it is then always possible to group ions into crystals such that:
\begin{equation}
\{B_C, B_2,\ldots,B_{m/2}\}\{B_{m/2+1},\ldots,B_m\}.
\end{equation}
In that case, the same reasoning as that for the case in Eq.~\ref{eqSys} applies, and the circuit fit is analogously bound by the ratio of the constant cost of an exchange over the number of gates in blocks $B_u$ of subset $H'$. Now, define $T_u$ as the number of gates in a given block $B_u$. As the number of gates varies between various blocks $B_u$, and assuming $T_{min} \leq T_u \leq T_{max}$ over the complete set of blocks $B_u$ of $H'$, the circuit fit is then bounded by
\begin{equation}
L_- = \frac{\text{Cost~of~one~exchange}}{T_{max}}\leq C,
\end{equation}
and 
\begin{equation}
C \leq \frac{\text{Cost~of~one~exchange}}{T_{min}}= L_+.
\end{equation}
As a result, the lower limit $L_-$ is reached when $T_{max}$ is highest, that is the circuit fit is improved with increased crystal size.


\section{Conclusion and outlook}

An algorithm for the generation of shuttling  schedules was successfully designed and implemented. We have shown that it generates efficient, close to optimal solutions for two important classes of quantum circuits. Future steps will extend this layer of the compilation stack to handle larger ($>2$) sizes of stored ion crystals, allow the shuttling compilation stage to handle advanced trap features such as junctions, and to investigate in general how capabilities and constraints imposed by a given trap architecture impact the incurred shuttling overhead. Furthermore, an interlinkage with the preceding gate compilation stage is to be established, such that gates sequences leading to significantly reduced shuttling overhead can be generated. Finally, the compilation stack is to be supplemented with routines for gate error estimation and mitigation, specifically tailored for shuttling-based platforms.

\section*{Acknowledgements}
J.D. acknowledges financial support from the QSciTech training program funded by the Collaborative Research and Training Experience (CREATE) program of the Natural Sciences and Engineering Research Council of Canada (NSERC).
Y.B.L. acknowledges support from the Canada First Research Excellence Fund (CFREF) through the Institut quantique at Université de Sherbrooke.
U.P. and F.S.K. acknowledge funding by the Germany ministry of science and education (BMBF) via the VDI within the projects HFAK and IQuAn. The research is based upon work supported by the Office of the Director of National Intelligence (ODNI), Intelligence Advanced Research Projects Activity (IARPA), via the U.S. Army Research Office grant W911NF-16-1-0070. The views and conclusions contained herein are those of the authors and should not be interpreted as necessarily representing the official policies or endorsements, either expressed or implied, of the ODNI, IARPA, or the U.S. Government. The U.S. Government is authorized to reproduce and distribute reprints for Governmental purposes notwithstanding any copyright annotation thereon. Any opinions, findings, and conclusions or recommendations expressed in this material are those of the author(s) and do not necessarily reflect the view of the U.S. Army Research Office.

\bibliographystyle{quantum}
\bibliography{citquat}

\end{document}